\documentclass{aa}
\usepackage{graphics,epsfig,lscape,times}

\usepackage{natbib}
\bibpunct{(}{)}{;}{a}{}{,}

\newcommand{\ergcm}[1]{$\times 10^{#1}$ erg cm$^{-2}$ s$^{-1}$}

\newcommand{\hcm}[1]{$\times 10^{#1}$ cm$^{-2}$}
\newcommand{\ohcm}[1]{$10^{#1}$ cm$^{-2}$}
\newcommand{\expo}[1]{$\times 10^{#1}$}
\newcommand{\oexpo}[1]{$10^{#1}$}
\newcommand{\nh}{N$_{\rm H}$}

\newcommand{\eline}{\hbox{E$_{\rm line}$}}
\newcommand{\wline}{\hbox{$\sigma_{\rm line}$}}
\newcommand{\eqw}{\hbox{EW}}

\newcommand{\ct}{cts s$^{-1}$}

\newcommand{\ltsima}{$\buildrel < \over \sim$}
\newcommand{\lsim}{\lower.5ex\hbox{\ltsima}}
\newcommand{\gtsima}{$\buildrel > \over \sim$}
\newcommand{\gsim}{\lower.5ex\hbox{\gtsima}}

\newcommand{\rxb}{\hbox{\object{RX\,J0720.4$-$3125}}}
\newcommand{\rxc}{\hbox{\object{RX\,J0806.4$-$4123}}}

\newcommand{\rxe}{\hbox{\object{RX\,J1605.3+3249}}}
\newcommand{\rxf}{\hbox{\object{RX\,J1856.4$-$3754}}}
\newcommand{\rbs}{\hbox{\object{RBS1223}}}

\begin{document}
 
\title{A phase-dependent absorption line in the spectrum of the X-ray pulsar \rxb
        \thanks{Based on observations with XMM-Newton,
               an ESA Science Mission with instruments and contributions 
               directly funded by ESA Member states and the USA (NASA)}}
 
\author{F.~Haberl \and V.E.~Zavlin \and J.~Tr{\"u}mper \and V.~Burwitz }

\titlerunning{A phase-dependent absorption line in the spectrum of the X-ray pulsar \rxb}
\authorrunning{Haberl et al.}
 
\offprints{F. Haberl, \email{fwh@mpe.mpg.de}}
 
\institute{Max-Planck-Institut f\"ur extraterrestrische Physik,
           Giessenbachstra{\ss}e, 85748 Garching, Germany}
 
\date{Received 29 July 2003/ Accepted 16 March 2004}
 
\abstract{
The XMM-Newton spectra of the isolated neutron star \rxb\ show deviations from a 
Planckian energy distribution below 400 eV, similar to the spectra of RBS1223, another
long-period X-ray pulsar, as reported recently by \citet{2003A&A...403L..19H}. 
For a Gaussian-shaped absorption line we derive an energy of 271 eV and 
an equivalent width of $-$40 eV from the phase-averaged spectra of \rxb. 
We investigate the spectral variations seen in hardness ratios as function of 
pulse phase and find that they are best described by changes in the depth of the
absorption line. The line equivalent width changes between $-$31 eV 
around intensity maximum of the pulse and $-$58 eV at the declining part of the pulse.
Small variations ($<$20 eV) of the line energy with pulse phase may still be caused 
by statistical fluctuations. On the other hand, the black-body temperature varies 
significantly by 2.5 eV (statistical 90\% errors typically 0.7 eV) reaching the highest value 
at pulse maximum.
One possible interpretation for the absorption line is cyclotron resonance scattering
of protons in a magnetic field with B $\simeq$ 5\expo{13} G. This field strength is 
compatible with estimates inferred from recent spin down measurements of the pulsar.

}
 
\maketitle
 
\section{Introduction}

The soft X-ray source \rxb\ was discovered in the ROSAT all-sky survey data by
\citet{1997A&A...326..662H}. The detection of 8.39 s sinusoidal pulsations 
in the X-ray flux, its black-body-like spectrum with little interstellar 
absorption and the identification with a faint blue optical star 
\citep{1998A&A...333L..59M,1998ApJ...507L..49K,2003ApJ...590.1008K,2003A&A...408..323M} classified 
this object as a nearby isolated neutron star. It is the second brightest of a 
group of seven, maybe 
eight, radio quiet \citep[recently stringent limits were obtained for
\rxb\ and \rxc,][]{2003MNRAS.340L..43J} and X-ray dim isolated neutron 
stars \citep[XDINs; for a recent review see ][]{haberl2002COSPAR}.

The available low-resolution X-ray spectra obtained by ROSAT from 
XDINs were all consistent with Planckian energy distributions of
temperatures kT in the range 40 -- 100 eV and little attenuation by 
interstellar absorption. More recent observations of \rxb\ and 
the brightest object \rxf\ were performed using the low
energy transmission grating spectrometer (LETGS) aboard Chandra and the reflection grating
spectrometers (RGSs) of XMM-Newton. Also at high spectral resolution both 
sources show no significant narrow features in their spectra which can best be 
modeled by a Planckian spectrum \citep{2001A&A...365L.298P,2001A&A...379L..35B}
or a modified Planckian spectrum \citep{2003A&A...399.1109B}. In particular 
the high signal to noise LETGS spectrum from a 500 ks Chandra observation of 
\rxf\ did not reveal any significant deviation from a smooth continuum spectrum
\citep{2003A&A...399.1109B}. The statistical quality and energy band coverage 
of the LETGS and RGS spectra of \rxb\ are, however, insufficient to detect broad and
shallow features in the spectrum.

Cyclotron resonance absorption features in the 0.1--1.0 keV band can be expected 
in spectra from magnetized neutron stars with field strengths in the range 
of \oexpo{10}--\oexpo{11} G or 2\expo{13}--2\expo{14} G if caused by electrons 
or protons, respectively.
The X-ray pulsar \rbs\ was the first XDIN for which a significant narrow-band
deviation (at $\sim$300 eV, equivalent width --150 eV, width $\sim$100 eV) from a 
Planckian spectrum was discovered \citep[][ hereafter H03]{2003A&A...403L..19H}.
This absorption feature was discussed in terms of proton cyclotron resonance
scattering in a magnetic field of (2--6)\expo{13} G.

H03 also suggested
that changes in the soft part of the X-ray spectrum of the pulsar \rxb\ with 
pulse phase reported by \citet[][ hereafter C01]{2001A&A...365L.302C}, are 
caused by variable cyclotron absorption.
In this paper we present the results from a spectral analysis of XMM-Newton data
on the isolated neutron star \rxb. The observed spectra show a broad absorption 
feature which varies with pulse phase. We discuss this in the framework
of proton cyclotron resonance or atomic line transitions in a strong magnetic field.

\begin{table}
\caption[]{XMM-Newton EPIC observations of \rxb.}
\begin{tabular}{cccccc}
\hline\noalign{\smallskip}
\multicolumn{2}{c}{Time (UT)} &
\multicolumn{1}{l}{XMM} &
\multicolumn{1}{c}{Read-out} &
\multicolumn{1}{c}{Count rate} &
\multicolumn{1}{c}{Exp.} \\
\multicolumn{1}{c}{Start} &
\multicolumn{1}{c}{End} &
\multicolumn{1}{l}{Detector} &
\multicolumn{1}{c}{Mode$^{(1)}$} &
\multicolumn{1}{c}{[s$^{-1}$]$^{(2)}$} &
\multicolumn{1}{c}{[ks]} \\

\noalign{\smallskip}\hline\noalign{\smallskip}
\multicolumn{6}{l}{2000 May 13, Satellite Revolution 078, Obs. ID 0124100101} \\
\noalign{\smallskip}\hline\noalign{\smallskip}
  02:43 & 19:59 & MOS1/2  & FF/SW & 1.63/1.87 & 62.0 \\
  02:30 & 20:00 & pn	  & FF    &    7.59   & 62.5 \\
  01:42 & 15:36 & RGS1/2  & SQ    & 0.26/0.23 & 50.0 \\   
\noalign{\smallskip}\hline\noalign{\smallskip}
\multicolumn{6}{l}{2000 Nov. 21-22, Satellite Revolution 175, 0132520301} \\
\noalign{\smallskip}\hline\noalign{\smallskip}
  21:18 & 02:21 & MOS1/2  & LW    & 1.31/1.40 & 18.2 \\
  19:21 & 02:36 & pn	  & FF    &    6.51   & 26.1 \\
  18:30 & 03:06 & RGS1/2  & SQ    & 0.24/0.23 & 30.9 \\   
\noalign{\smallskip}\hline\noalign{\smallskip}
\multicolumn{6}{l}{2002 Nov. 6-7, Satellite Revolution 533, 0156960201} \\
\noalign{\smallskip}\hline\noalign{\smallskip}
  17:52 & 02:14 & MOS1/2  & FF    &    --     & 30.0 \\
  18:14 & 02:15 & pn	  & FF    &    7.66   & 28.4 \\
  17:51 & 02:16 & RGS1/2  & SQ    & 0.23/0.22 & 30.2 \\   
\noalign{\smallskip}\hline\noalign{\smallskip}
\multicolumn{6}{l}{2002 Nov. 8-9, Satellite Revolution 534, 0156960401} \\
\noalign{\smallskip}\hline\noalign{\smallskip}
  19:25 & 04:18 & MOS1/2  & FF    &    --     & 31.8 \\
  19:47 & 04:18 & pn	  & FF    &    7.64   & 30.2 \\
  19:24 & 04:19 & RGS1/2  & SQ    & 0.23/0.22 & 32.0 \\   
\noalign{\smallskip}
\hline
\end{tabular}

$^{(1)}$ FF: Full Frame (time resolution of 0.073 s and 2.6 s for EPIC-pn and -MOS, respectively; 
         SW: Small Window (MOS time resolution 0.3 s); LW: Large Window (MOS time resolution 0.9 s); SQ: Spectro+Q\\
$^{(2)}$ Mean count rates are for the energy bands 0.13--1.5 keV (EPIC) and 
         0.32--0.90 keV (RGS); when not given the data are not used in the spectral analysis.\\
The thin filter was used in all EPIC cameras in revolutions 78, 533 and 534 
and the medium filter in revolution 175.
\label{xmm-obs}
\end{table}

\section{XMM-Newton observations}

The X-ray pulsar \rxb\ was observed with XMM-Newton \citep{2001A&A...365L...1J} on five
occasions. Here we utilize the data collected with the European Photon Imaging 
Cameras (EPICs) which provide medium energy-resolution X-ray spectra. The cameras 
are based on MOS \citep[EPIC-MOS1 and \hbox{-MOS2,}][]{2001A&A...365L..27T} and 
pn \citep[EPIC-pn,][]{2001A&A...365L..18S} CCD technology and are mounted in the 
focal plane of the three X-ray multi-mirror systems \citep{2000SPIE.4012..731A}.
Simultaneous high spectral resolution data were obtained with the Reflection 
Grating Spectrometers \citep[][]{2001A&A...365L...7D}.
The details of the XMM-Newton observations are summarized in Table~\ref{xmm-obs}.
We do not use the last observation because EPIC-pn was operated in SW
read-out mode (less well calibrated at low energies) and with 
the thick filter. During the first observation the EPIC instruments suffered from 
strong background flares which were screened out leaving only 36 ks (pn) and 
39 ks (MOS) which were used for the spectral analysis. From the other three 
observations all the data was used.

In processing and preparation of the observational data and model fitting of the spectra 
we followed H03 (using an extraction radius of 40\arcsec\ for the EPIC data). 
We used the same version of the XMM-Newton analysis software SAS and calibration files.

\section{Pulse-phase averaged spectra}

\subsection{Combined XMM-Newton spectra}

\begin{figure*}
\begin{center}
\begin{minipage}[t]{8.6cm}
\resizebox{8.6cm}{!}{\includegraphics[clip=]{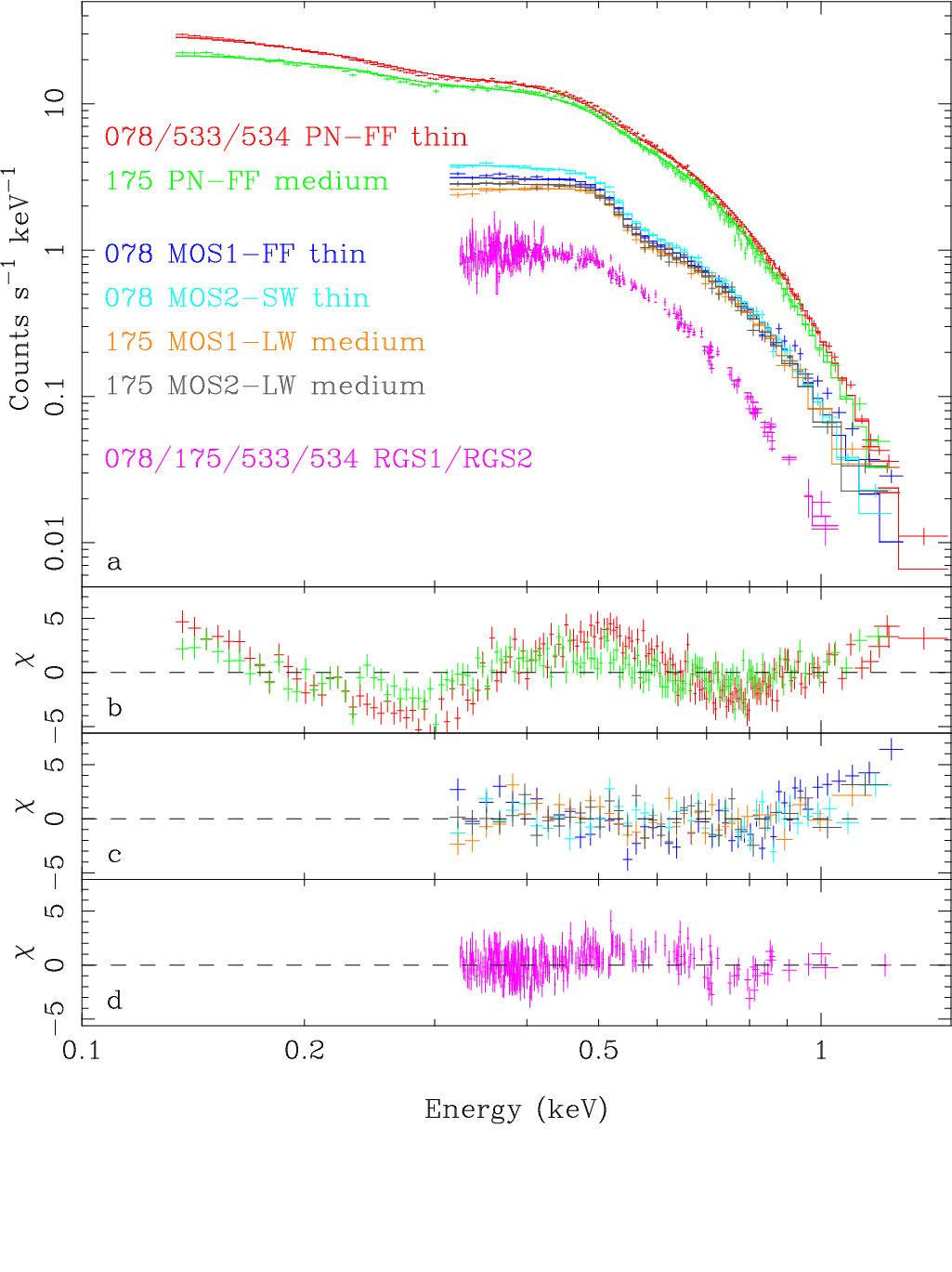}}
\end{minipage}
\hspace{2mm}
\begin{minipage}[t]{8.6cm}
\resizebox{8.6cm}{!}{\includegraphics[clip=]{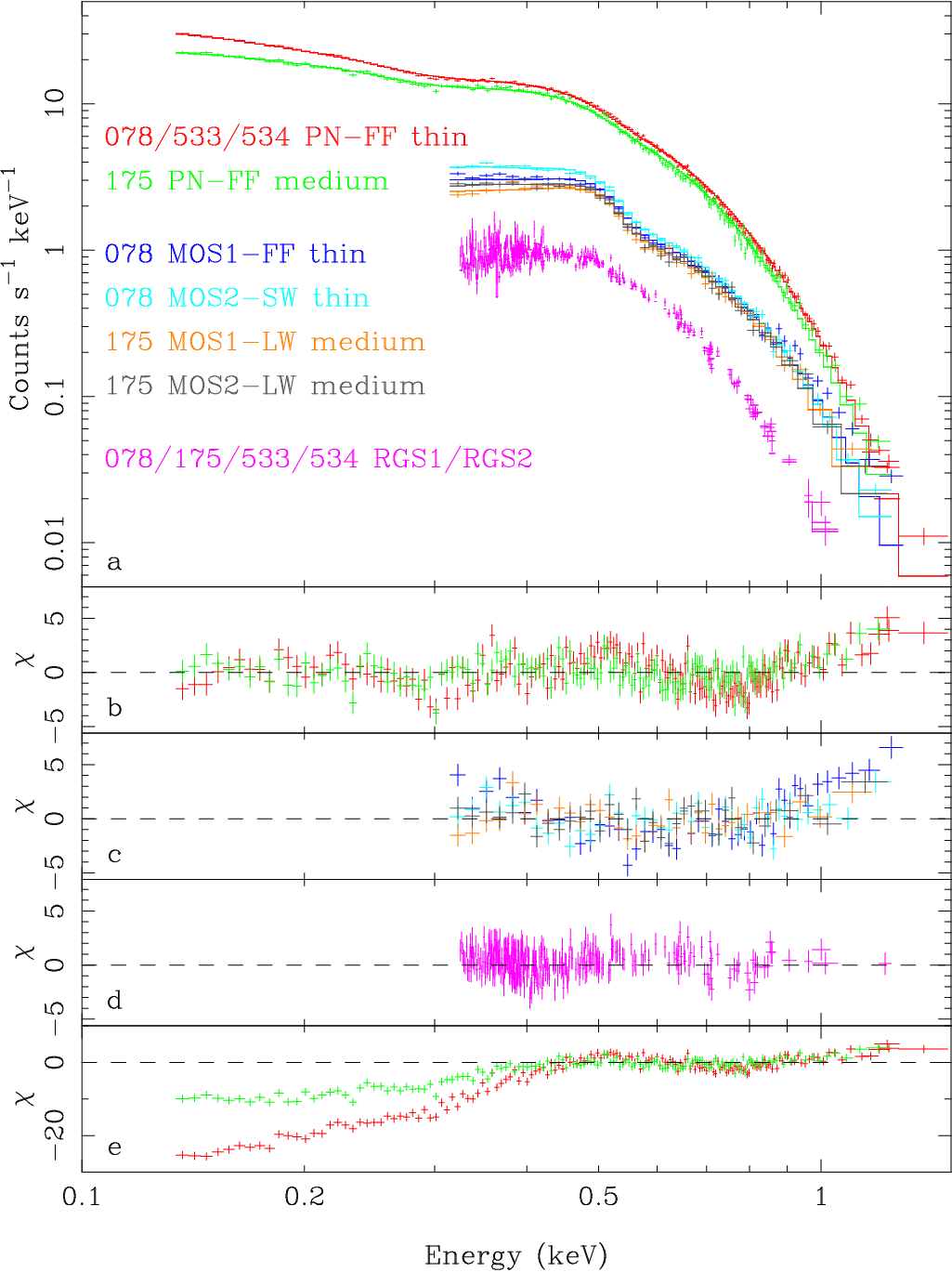}}
\end{minipage}
\end{center}
\caption{Simultaneous fits using models A (left) and B (right) to the XMM-Newton 
spectra of \rxb. For model definition see Table~\ref{tab-fits}.
For each model the best fit (histogram) to the spectra (crosses) is plotted in panels (a).
Panels (b) -- (d) show the residuals for EPIC-pn, -MOS and RGS spectra, respectively.
For model B panel (e) illustrates the best fit model with the absorption line removed.
The three EPIC-pn spectra obtained with thin filter were combined for clarity in the plots, 
as well as all the eight RGS spectra. The MOS data below 300 eV were not used for the 
spectral fits. 
The residuals increasing with energy above 800 eV in the EPIC spectra are probably 
caused by pile-up (see Sect.~\ref{sect-pileup}).}
\label{fig-spectra}
\end{figure*}

H03 reported the analysis of the XMM-Newton spectra of \rbs. They 
could not find satisfactory fits using a simple black-body model, non-magnetic neutron 
star atmosphere models with different element abundances or two black-body components 
with different temperatures. An acceptable fit was found with a broad absorption 
line included in the black-body model. In the following analysis of the spectra of 
\rxb\ we find very similar residuals when comparing the measured spectra to the 
simple absorbed black-body model.
We therefore also include a Gaussian-shaped absorption line in the modeling of the 
spectra of \rxb. To account for cross-calibration problems between the different 
instruments we allow the black-body temperature to be fit individually to each 
spectrum but fit the absorption column density as single free parameter to all 
sixteen XMM-Newton spectra simultaneously. Similarly, all three parameters of 
the absorption line (energy, width and normalization) were fit to all spectra 
simultaneously. The results for both models are reported in Table~\ref{tab-fits}.
The line at 271 eV is relatively broad with a width of $\sigma$ = 64 eV. The line equivalent 
width (\eqw) is about $-$40 eV\footnote{All values for \eqw\ are obtained with the spectral 
fitting package XSPEC version 11.2.0.}, less prominent as measured in the spectra of \rbs\ with 
\eqw\ = $-$150 eV (H03).
The XMM-Newton spectra of \rxb\ together with the best fit models are 
shown in Fig.~\ref{fig-spectra}.

\begin{figure}
\begin{center}
\resizebox{8.6cm}{!}{\includegraphics[clip=,angle=-90]{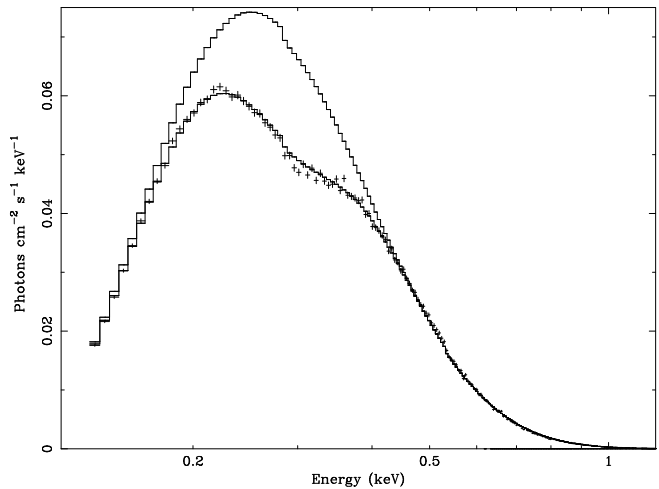}}
\end{center}
\caption{Incident (`unfolded') photon spectrum for the black-body model with absorption line. 
The crosses represent the combined EPIC-pn spectrum from the three thin-filter observations.
The best fit black-body model with absorption line and with the line normalization 
set to zero are plotted as solid and dashed histograms, respectively.}
\label{fig-unfolded}
\end{figure}

The deviation from a Planckian shape is highly significant as can be seen from the
improvement in the $\chi^2$ values (Table~\ref{tab-fits}) 
when an absorption line is included ($\Delta\chi^2$=683 for 
3 additional parameters in the fits to the spectra of the EPIC-pn instrument which is 
most sensitive at energies below 300 eV). 
In Fig.~\ref{fig-unfolded} the incident photon spectrum is shown (for clarity only the summed
EPIC-pn spectrum from the three thin filter observations is plotted). This 
presentation better visualizes the absorption feature as compared to 
Fig.~\ref{fig-spectra} where the detector response at energies below 500 eV 
(with relatively bad energy resolution and strong energy redistribution into lower detector
channels due to
electron losses in the surface layers of the detector) smears out the intrinsically 
broad feature.

\begin{table*}[t]
\caption[]{Spectral analysis of pulse-phase averaged X-ray spectra of \rxb.}
\begin{tabular}{l|ccc|ccccccc}
\hline\noalign{\smallskip}
\multicolumn{1}{l|}{} &
\multicolumn{3}{c|}{Model A: phabs*bbody$^{(1)}$} &
\multicolumn{7}{c}{Model B: phabs*(bbody+gaussian)$^{(1)}$} \\

\multicolumn{1}{l|}{Obs-Inst.} &
\multicolumn{1}{c}{kT} &
\multicolumn{1}{c}{\nh} &
\multicolumn{1}{c|}{$\chi^2$/dof} &
\multicolumn{1}{c}{kT} &
\multicolumn{1}{c}{\nh} &
\multicolumn{1}{c}{\eline} &
\multicolumn{1}{c}{\wline} &
\multicolumn{1}{c}{\eqw$^{(2)}$} &
\multicolumn{1}{c}{$\chi^2$/dof} &
\multicolumn{1}{c}{Flux$^{(3)}$} \\

\multicolumn{1}{l|}{} &
\multicolumn{1}{c}{[eV]} &
\multicolumn{1}{c}{[\oexpo{20}cm$^{-2}$]} &
\multicolumn{1}{c|}{per inst.} &
\multicolumn{1}{c}{[eV]} &
\multicolumn{1}{c}{[\oexpo{20}cm$^{-2}$]} &
\multicolumn{1}{c}{[eV]} &
\multicolumn{1}{c}{[eV]} &
\multicolumn{1}{c}{[eV]} &
\multicolumn{1}{c}{per inst.} &
\multicolumn{1}{c}{[erg cm$^{-2}$ s$^{-1}$]} \\

\noalign{\smallskip}\hline\noalign{\smallskip}
078-pn   &  85.2 & 1.38      &           & 84.3$\pm$0.3 & 0.79$\pm$0.13 & 271$\pm$14 & 64$\pm$7   & $-$39.6 &           & 1.19\expo{-11}\\
175-pn   &  84.6 & =$^{(4)}$ &           & 83.4$\pm$0.4 & =  	        & =	     &  =         & $-$39.4 & 	      & 1.22\expo{-11}\\
533-pn   &  88.6 & =	     &	         & 87.5$\pm$0.4 & =  	        & =	     &  =         & $-$40.6 & 	      & 1.19\expo{-11}\\
534-pn   &  88.3 & =	     & 1679/639  & 87.2$\pm$0.4 & =  	        & =	     &  =         & $-$40.5 &  996/636  & 1.19\expo{-11}\\
078-MOS1 &  84.4 & =         &	         & 84.1$\pm$0.6 & =  	        & =	     &  =         & $-$39.6 & 	      & 1.21\expo{-11}\\
175-MOS1 &  83.8 & =	     &	         & 83.5$\pm$0.8 & =  	        & =	     &  =         & $-$39.4 & 	      & 1.16\expo{-11}\\
078-MOS2 &  81.6 & =	     &	         & 81.3$\pm$0.6 & =  	        & =	     &  =         & $-$38.8 & 	      & 1.49\expo{-11}\\
175-MOS2 &  82.7 & =	     & 408/167   & 82.4$\pm$0.8 & =             & =	     &  =         & $-$39.1 & 468/167   & 1.28\expo{-11}\\
078-RGS1 &  86.7 & =         &           & 85.0$\pm$1.1 & =  	        & =	     &  =         & $-$39.8 & 	      & 0.99\expo{-11}\\
175-RGS1 &  89.6 & =	     &	         & 87.7$\pm$1.5 & =  	        & =	     &  =         & $-$40.7 & 	      & 0.94\expo{-11}\\
533-RGS1 &  95.0 & =	     &	         & 93.1$\pm$1.6 & =  	        & =	     &  =         & $-$42.7 & 	      & 0.86\expo{-11}\\
534-RGS1 &  94.1 & =         &           & 92.2$\pm$1.6 & =  	        & =	     &  =         & $-$42.4 & 	      & 0.88\expo{-11}\\
078-RGS2 &  86.7 & =         &	         & 85.3$\pm$1.1 & =  	        & =	     &  =         & $-$39.9 & 	      & 0.96\expo{-11}\\
175-RGS2 &  88.4 & =         &	         & 87.0$\pm$1.4 & =  	        & =	     &  =         & $-$40.5 & 	      & 0.92\expo{-11}\\
533-RGS2 &  92.8 & =         &	         & 91.5$\pm$1.4 & =  	        & =	     &  =         & $-$42.1 & 	      & 0.85\expo{-11}\\
534-RGS2 &  94.4 & =         & 1453/1266 & 92.9$\pm$1.4 & =  	        & =	     &  =         & $-$42.6 & 1455/1266 & 0.85\expo{-11}\\
total    &       &           & 3540/2072 &              &               &            &            &       & 2919/2069 & \\
\noalign{\smallskip}\hline\noalign{\smallskip}
LETGS    &  82.5 & 1.50      & 157/181   & 78.8$\pm$2.0 & 1.56$\pm$0.20 & 271f 	     &  64f       & --    & 157/181   & 0.99\expo{-11}\\
\noalign{\smallskip}\hline\noalign{\smallskip}
\end{tabular}

All errors are given for a 90\% confidence level. Parameters fixed in the fit are marked by ``f"
behind the value.\\
$^{(1)}$ Standard XSPEC names for the model components.\\
$^{(2)}$ For MOS and RGS spectra the line \eqw\ is calculated from the 
         extrapolated model because of the restricted line coverage.\\
$^{(3)}$ Observed flux 0.1--2.4 keV.\\
$^{(4)}$ ``=" denotes fit parameter is forced to be the same for all spectra.
\label{tab-fits}
\end{table*}

\subsection{A problem in the spectral calibration?}
\label{sect-pncal}

\citet{2001A&A...365L.298P} presented a black-body fit to the EPIC-pn spectrum of \rxb\ 
obtained from the observation in May 2000. Residuals around 300 eV were attributed to uncertainties
in the filter transmission data. However, the modeling was based on early spectral response
calibration data and \rxb\ was the only XDIN observed with XMM-Newton at that time. In addition to \rxb\
and \rxf\ four XDINS are now observed with XMM-Newton, partly several times and different 
instrumental setup and we have access to all the data. The spectral modeling of 
the spectra is in progress and revealed significant deviations
from a Planckian shape for \rbs\ (H03), \rxb\ (as presented here) and \rxe\ \citep{2004vanKerkwijk}. 

\begin{figure}
\begin{center}
\resizebox{8.6cm}{!}{\includegraphics[clip=,angle=-90]{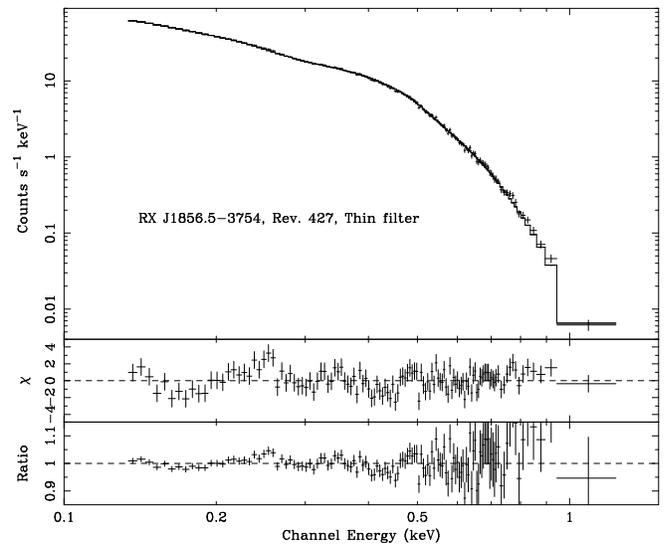}}
\end{center}
\caption{Black-body model fit to the EPIC-pn spectrum of \rxf.
The lower two panels show the residuals from the best fit in units of $\sigma$ and as ratio
data/model.}
\label{fig-pncal}
\end{figure}

The best quality X-ray spectrum of any XDIN was obtained from \rxf\ with the Chandra LETGS. It is 
well described by a Planckian model\footnote{As discussed in \citet{2003A&A...399.1109B} the broad-band 
spectrum of \rxf\ is even better modeled by a modified black-body spectrum of the form 
$E^\beta \times B_\nu$ ($B_\nu$ is the Planck function). We note that the spectral deviations of \rxb\ 
discussed in this paper are found in a relatively narrow energy band in contrast to the smooth wide-band
modification caused by the power-law component $E^\beta$ which mainly changes the black-body parameters.} 
and can be used as a calibration reference for the soft 
X-ray band. Black-body fits to the EPIC-pn spectrum of \rxf, using the same instrument calibration 
as for \rxb\ are fully acceptable, with deviations less than 5\% at all energies below 
1 keV\footnote{See the presentation by M. Kirsch at\\
http://xmm.vilspa.esa.es/external/xmm\_user\_support/usersgroup/ 20030331/presentations.shtml}.
For clarity and better comparison with \rxb\ we show the EPIC-pn spectrum of \rxf\
with best black-body fit in Fig.~\ref{fig-pncal}. 
The residuals in the black-body fit to the spectra of \rxb\ (Fig.~\ref{fig-spectra}) 
clearly show a strong systematic behavior, 
which is not seen in the case of \rxf. Hence, we conclude that the deviations from a Planckian shape 
revealed in the EPIC-pn spectra of \rxb\ are not caused by calibration uncertainties.
In addition, the strength of the observed feature varies with pulse phase (see Sect. 4).

However, as can be seen from Table~\ref{tab-fits}, there are still cross-calibration problems 
between the different instruments which cause systematic uncertainties larger than
the statistical errors. 
The EPIC-MOS spectra are barely consistent to each other at low energies and there seems to be
a systematic difference between the residuals of the model fit for thin and medium filter 
spectra (medium filter spectra better agree with those of EPIC-pn down to 200 eV). We excluded 
therefore the MOS data below 300 eV in the spectral analysis. The EPIC-pn spectra are 
in large agreement with each other, although those obtained from the last two observations
yield significantly higher black-body temperatures. 
These temperature differences are caused by apparent gain variations by up to 3\% at 
energies below 1 keV which were
measured in emission line spectra of supernova remnants (routine calibration targets to monitor
the energy scale) and the instrumental Oxygen K-edge energy determined from strong sources
with flat continuum spectra\footnote{See the presentation by F. Haberl at\\ 
ftp://epic3.xra.le.ac.uk/pub/cal-pv/meetings/tuebingen-2003-02}.
The origin of this effect is not well understood and can presently not be corrected for. Gain 
variations shift the spectral part above 500 eV
which results in the temperature change, while the spectrum below 500 eV, which
determines the \nh\ (and where the deviations from a black-body model are seen),
is practically unaffected (allowing \nh\ to vary in the EPIC-spectra yields
values consistent within the errors). To account for the gain variations we therefore allowed
kT to vary between the spectra.
The somewhat higher flux\footnote{Errors on the pn fluxes are 7\ergcm{-14}} derived from the 
revolution 175 spectrum may be caused by reduced pile-up losses (see Sect.~\ref{sect-pileup}). 
\subsection{Pile-up in the EPIC spectra}
\label{sect-pileup}

Bright sources observed with X-ray CCD detectors can be subject to pile-up effects.
If the read-out is not sufficiently fast, two or more photons may deposit charge
in one or few neighboring CCD pixels during the exposure of a single read-out frame. The energy of 
the detected event will then be the sum of the energies of the original photons. 
Therefore pile-up hardens a continuum spectrum.  
\rxb\ was observed by the EPIC-pn in full frame mode which provides pile-up free 
spectra for count rates below $\sim$6 \ct. 
As can be seen from the count rates listed in Table~\ref{xmm-obs} the EPIC-pn spectra 
are somewhat affected by pile-up. This probably explains the residuals
in the model fits at energies above $\sim$800 eV (Fig.~\ref{fig-spectra}).
Also the EPIC-pn spectrum obtained with medium filter (which reduces the total count rate) 
shows the lowest pn-derived black-body temperature and the highest flux (lowest flux 
losses).

Pile-up occurs
predominantly in the center of the point spread function (PSF) where the 
probability is highest that two photons are recorded in the same pixel within 
a single read-out frame. The SAS task `epatplot' can be used to visualize 
the pile-up effects on the spectrum. The main effect on very soft spectra 
like those of XDINs is a spectral hardening which leads to a somewhat higher 
black-body temperature and to some flux-loss when using single-pixel events only. 
In the case of \rxb\ we excluded the inner area of the PSF with 20\arcsec\ 
in radius (15\arcsec\ for the medium filter observation with lower 
count rate) to obtain practically pile-up free spectra. The disadvantage of 
this procedure is in reduction of the number of extracted counts by a factor of 7.3 
(4.0 in the medium filter case). The statistics becomes almost insufficient to 
investigate the deviations from a black-body spectrum and therefore, we fixed
the line energy and width at the values derived in Table~\ref{tab-fits}.
Table~\ref{tab-nopileup-fits} summarizes the black-body model parameters
inferred from the pile-up free EPIC-pn spectra.
The derived temperatures are typically 3 eV lower than those obtained from the spectra
not corrected for pile-up effects. The column density derived for both cases 
is consistent within the errors. There is again a significant difference in temperature 
between the first and last two spectra. However, at the current status of the 
calibration it is not clear if this is a long-term systematic effect.

\begin{figure}
\begin{center}
\resizebox{8.6cm}{!}{\includegraphics[clip=,angle=-90]{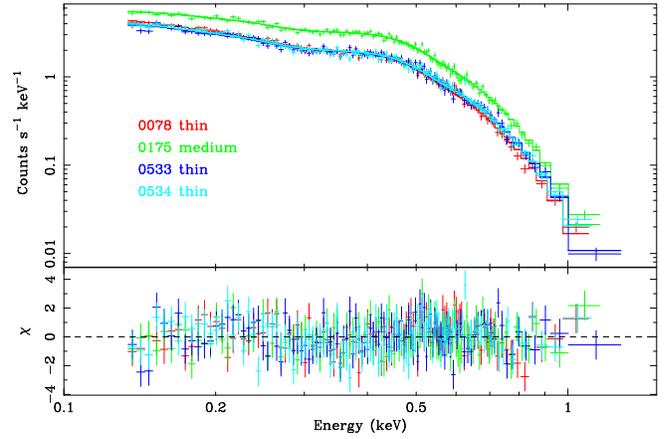}}
\end{center}
\caption{EPIC-pn spectra extracted from a ring-shaped area (inner and outer radii 
are 20\arcsec and 40\arcsec\ for the thin filter, and 15\arcsec and 40\arcsec\ for the medium 
filter observations), excluding the inner part of the PSF in order to avoid pile-up effects.}
\label{fig-nopileup-spectra}
\end{figure}

\begin{table}[t]
\caption[]{Fits to pile-up free EPIC-pn spectra.}
\begin{tabular}{lccccc}
\hline\noalign{\smallskip}
\multicolumn{4}{c}{Model B: phabs*(bbody+gaussian)} \\

\multicolumn{1}{l}{Obs-Inst.} &
\multicolumn{1}{c}{kT} &
\multicolumn{1}{c}{\nh} &
\multicolumn{1}{c}{\eline} &
\multicolumn{1}{c}{\wline} &
\multicolumn{1}{c}{$\chi^2$/dof} \\

\multicolumn{1}{l}{} &
\multicolumn{1}{c}{[eV]} &
\multicolumn{1}{c}{[\oexpo{20}cm$^{-2}$]} &
\multicolumn{1}{c}{[eV]} &
\multicolumn{1}{c}{[eV]} &
\multicolumn{1}{c}{per inst.} \\

\noalign{\smallskip}\hline\noalign{\smallskip}
078-pn & 81.6$\pm$0.9 & 0.90$\pm$0.13 & 271f & 64f &	     \\
175-pn & 81.6$\pm$0.9 & =	      &	=    & =   &         \\
533-pn & 85.2$\pm$1.0 & =	      &	=    & =   &         \\
534-pn & 84.7$\pm$1.0 & =	      & =    & =   & 477/439 \\
\noalign{\smallskip}\hline\noalign{\smallskip}
\end{tabular}
\label{tab-nopileup-fits}
\end{table}

\subsection{Comparison with other instruments}

\begin{figure}
\begin{center}
\resizebox{8.6cm}{!}{\includegraphics[clip=,angle=-90]{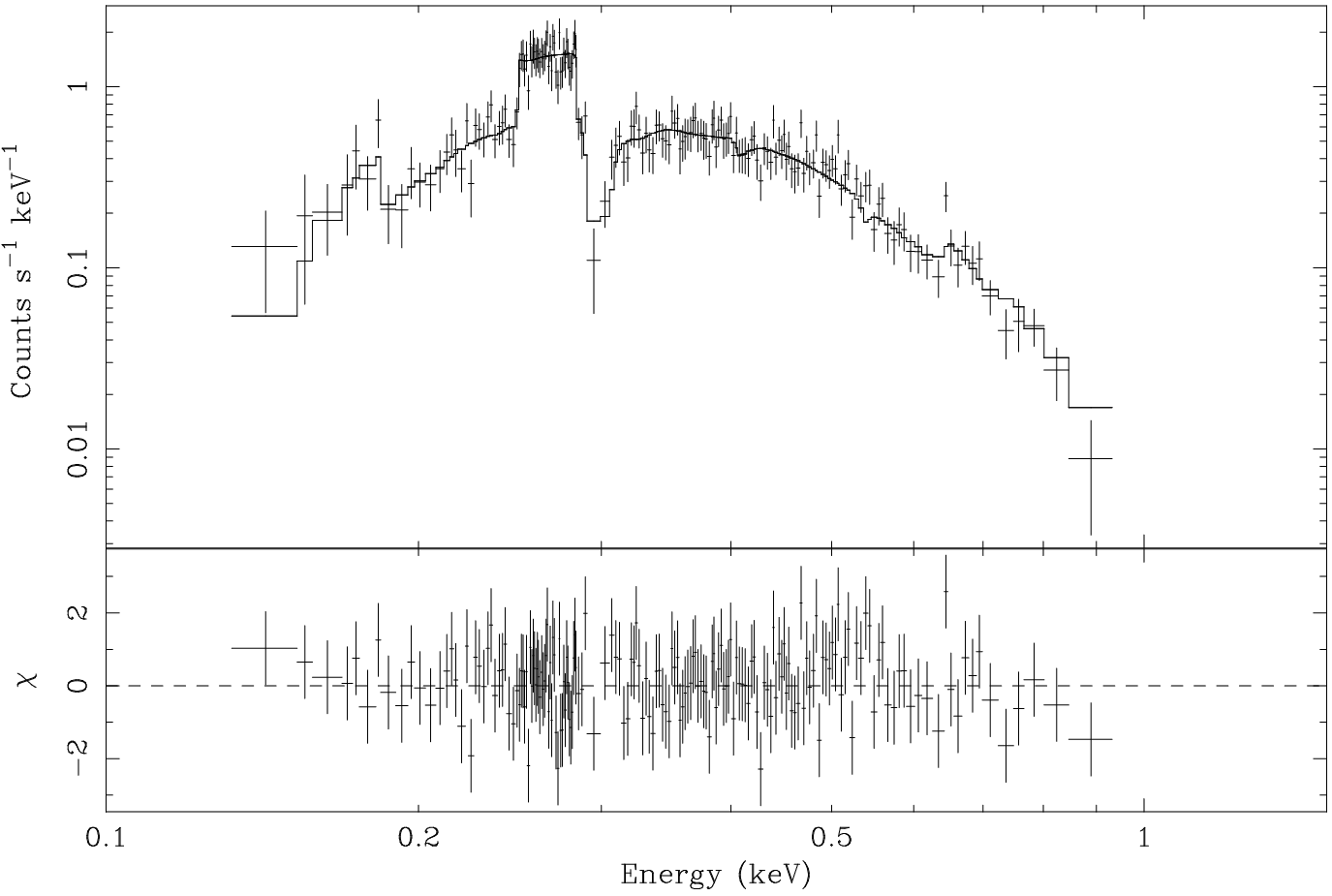}}
\resizebox{8.6cm}{!}{\includegraphics[clip=,angle=-90]{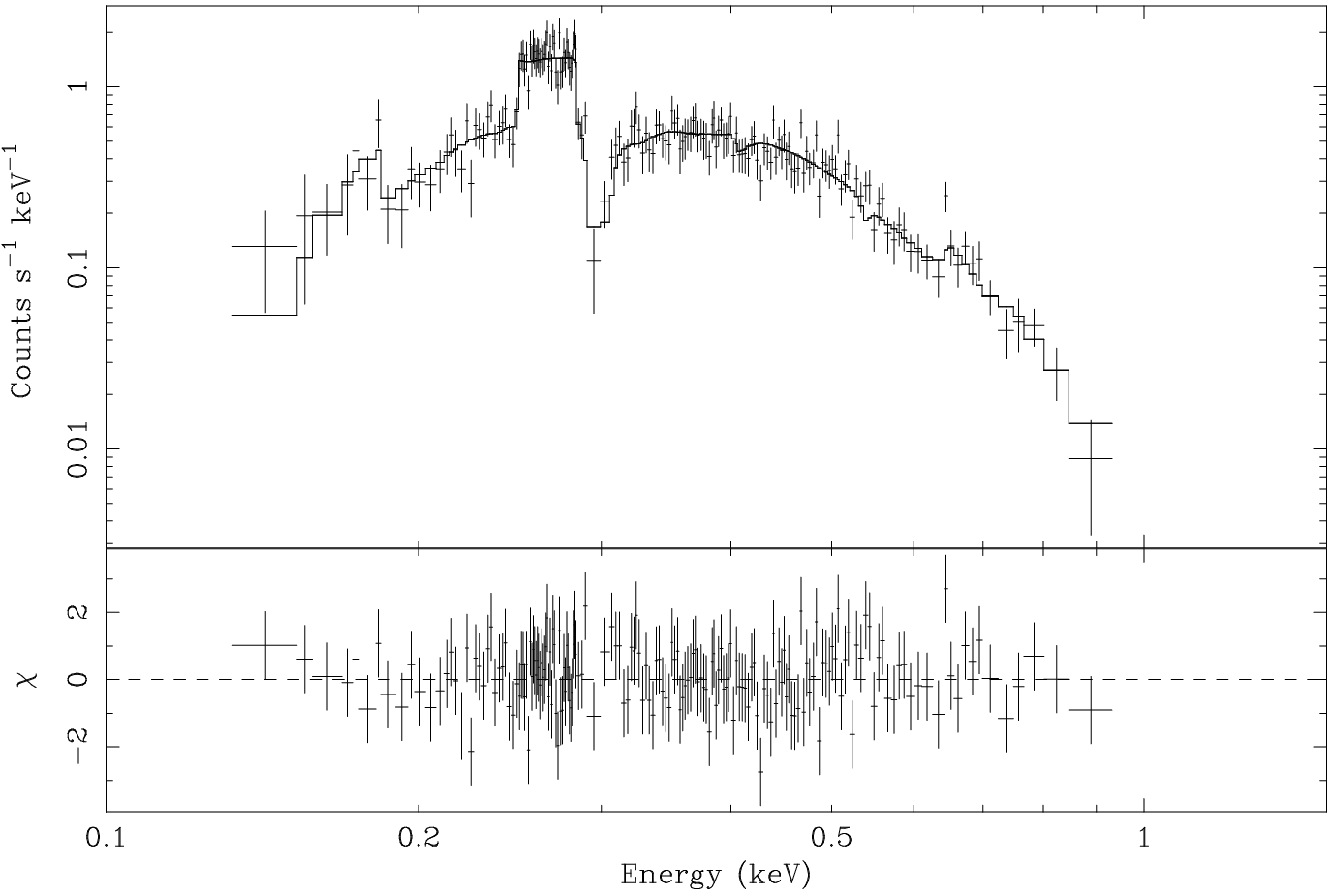}}
\end{center}
\caption{Fits using a black-body model with (bottom panels) and without absorption line 
to the LETGS spectra of \rxb. The low statistical quality of the spectrum
does not allow to discriminate between the models.}
\label{fig-letg}
\end{figure}

Chandra observed \rxb\ in February 2000 with the LETGS for a total of $\sim$38 ks 
(observation IDs 368, 369 and 745). The LETGS spectrum was found to be consistent 
with a black-body model with kT $\simeq$80 eV and absorption column density 
$\simeq$1.7\hcm{20} \citep{2002nsps.conf..273P}. We re-analyzed these data and 
obtained kT = 82.5$\pm$4.0 eV and \nh\ = (1.5$\pm$0.5)\hcm{20} for a single black-body model.
To verify that the LETGS spectrum is also consistent with the presence of an absorption
feature we added a Gaussian-shaped line, with parameters as derived from the
XMM-Newton spectra, to the black-body model. The fit results for both models,
with and without line, are listed in Table~\ref{tab-fits}. The statistical quality
of the LETGS spectrum is clearly insufficient to discriminate between the models
(Fig.~\ref{fig-letg}).
The column density inferred from the combined fit to the XMM-Newton spectra
is significantly lower than the LETGS value. Allowing individual values of \nh\ for the 
three XMM-Newton instruments (all other parameters in model B are treated as described in Table~\ref{tab-fits})
results in \nh/\ohcm{20} = 1.23$\pm$0.07, 0.66$\pm$0.30 and 
0.73$\pm$0.30 for EPIC-pn, -MOS and RGS, respectively, revealing still cross-calibration uncertainties.
The EPIC-pn value is in best agreement with the LETGS value which is probably because  
the LETGS spectrum of \rxf\ was used for calibration of the low-energy spectral response
of EPIC-pn.

\section{Pulse-phase variations}

\begin{figure}
\begin{center}
\resizebox{8.6cm}{!}{\includegraphics[clip=,angle=-90]{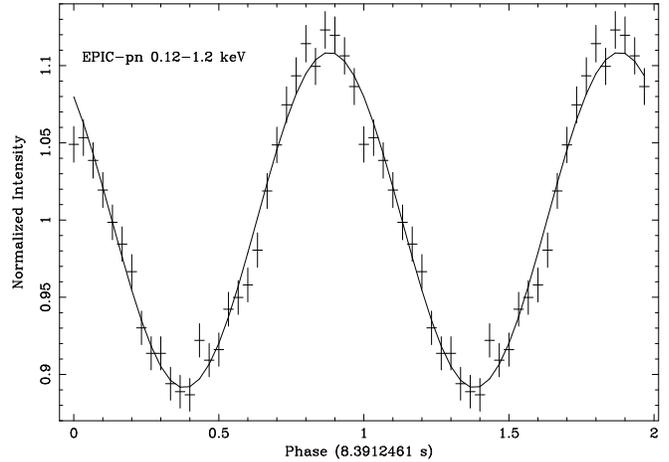}}
\end{center}
\caption{Pulse profile of \rxb\ obtained from the observation in revolution 534 together with
         the best fit sine wave.}
\label{fig-pulse}
\end{figure}

\begin{figure}
\begin{center}
\resizebox{8.6cm}{!}{\includegraphics[clip=,angle=-90]{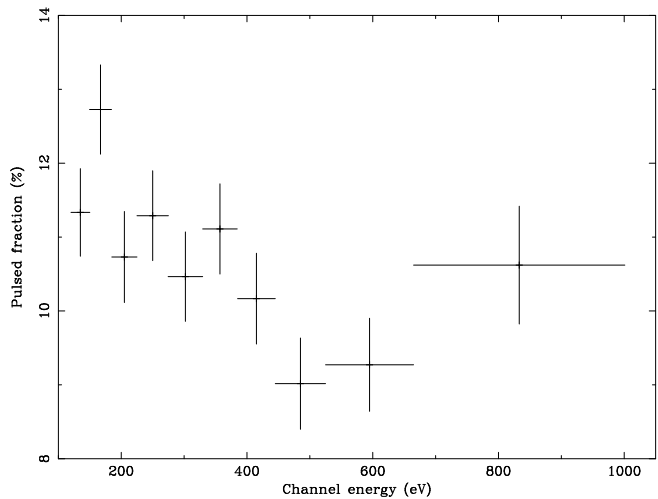}}
\end{center}
\caption{Pulsed fraction as function of energy obtained from the combined EPIC-pn thin filter data of 
         \rxb\ from revolutions 78, 533 and 534. For each energy bin a light curve with 10 phase bins
	 was folded and the pulsed fraction determined by (N$_{\rm t}$-N$_{\rm DC}$)/N$_{\rm t}$ with 
	 N$_{\rm t}$ beeing the total number of counts and N$_{\rm DC}$ the number of counts at the DC level
	 of the light curve. The energy bins were chosen to obtain at least 5000 events
	 for each phase bin.}
\label{fig-pfraction}
\end{figure}

\begin{figure}
\begin{center}
\resizebox{8.6cm}{!}{\includegraphics[clip=,angle=-90]{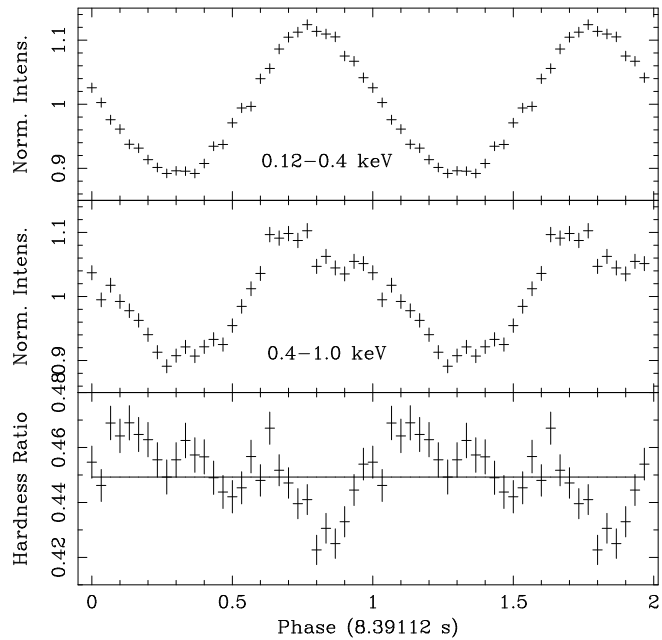}}
\end{center}
\caption{Pulse profile of \rxb\ obtained from the combined EPIC-pn thin filter data as used 
         in Fig.~\ref{fig-pfraction}
         in two energy bands, together with the hardness ratio. The line in the bottom panel
	 marks the best-fit with a constant hardness ratio which yields a $\chi^2_\nu$=4.6 for 29 dof.}
\label{fig-hardness}
\end{figure}

Because of the highest sensitivity, better time resolution (Table~\ref{xmm-obs}) and full
line coverage we investigate only EPIC-pn spectra in our pulse-phase spectroscopic study.
In the 0.12-1.2 keV energy band the pulse profile of \rxb\ is nearly sinusoidal. We fitted a 
sine wave to the folded light curves obtained from the EPIC-pn data in this band to determine
the pulsed fraction. An example is shown in Fig.~\ref{fig-pulse}. 
The pulsed fraction (after background correction) is around 11\%. Pulse periods and pulsed 
fractions inferred from the four EPIC-pn data sets are listed in Table~\ref{tab-pn-pf}. 
The presented pulse periods are used here to extract the phase-resolved spectra and we refer
for a more detailed temporal analysis to \citet{2004Cropper}.
The smaller pulsed fraction 
found in the revolution 175 data collected with medium filter (which more strongly
absorbs low-energy photons), suggests an energy dependence of the pulse profile. 
To investigate this further we determined the pulsed fraction as a function of energy
as plotted in Fig.~\ref{fig-pfraction}. We combined the data from the three thin filter observations
assuming a period change of 5\expo{-14} s s$^{-1}$ \citep{2002MNRAS.334..345Z} when folding the data.
There is an indication that the pulsed fraction
is larger below $\sim$400 eV compared to that at higher energies. We note here that due to the 
strong re-distribution at 
low energies the data points below 400 eV are not statistically independent (as discussed in Sect.~3.1
and demonstrated in panel ``e" of Fig.~\ref{fig-spectra}). Also, this approach is not sensitive to 
changes in the folded light curves related to the shape of the pulse profiles.
In the next step we therefore regard only two energy bands below and above 400 eV.
In Fig.~\ref{fig-hardness} the two light curves are plotted together with their ratio (hardness ratio). 
A dependence of the hardness ratio on pulse phase was first 
reported by C01. At energies above 400 eV the pulse peak becomes broader and more flat-topped. 
This results in a hardness ratio minimum near the intensity maximum and two (asymmetric) 
hardness ratio maxima at the flanks of the intensity peak. A $\chi^2$ test for constancy of
the hardness ratio yields a reduced value $\chi^2_\nu$ = 4.6 (for 29 degrees of freedom).
The corresponding probability that the hardness ratio is constant is extremely low: 2.5\expo{-15}.

\begin{table}
\caption[]{X-ray pulsations derived from the EPIC-pn data of \rxb.}
\begin{tabular}{ccc}
\hline\noalign{\smallskip}
\multicolumn{1}{l}{Rev.} &
\multicolumn{1}{c}{Pulse} &
\multicolumn{1}{c}{Pulsed$^{(1)}$} \\
\multicolumn{1}{l}{} &
\multicolumn{1}{c}{Period} &
\multicolumn{1}{c}{Fraction} \\
\multicolumn{1}{l}{} &
\multicolumn{1}{c}{[s]} &
\multicolumn{1}{c}{[\%]} \\

\noalign{\smallskip}\hline\noalign{\smallskip}
 078 & 8.3911165 & 10.5$\pm$0.3 \\
 175 & 8.3912190 & 10.1$\pm$0.4 \\
 533 & 8.3911276 & 11.1$\pm$0.4 \\
 534 & 8.3912461 & 11.1$\pm$0.4 \\
\noalign{\smallskip}
\hline
\end{tabular}

$^{(1)}$ 0.12--1.2 keV
\label{tab-pn-pf}
\end{table}

We divided the pulse into four
phase intervals of equal length in order to investigate the X-ray 
spectrum as function of pulse phase. The intervals cover the intensity minimum,
the rise, the intensity maximum and the decline of the pulse. The latter phase interval
is centered at the hardness ratio maximum derived from the EPIC-pn data. 
Compared to the approach above, spectral modeling has the advantage that the spectral response 
(energy resolution, re-distribution) is taken into account. However, for statistical reasons
only a small number of phase bins can be investigated.
The sixteen phase-selected spectra (from four observational data sets) were fit 
again simultaneously with the model
including an absorption line. Because of the reduced statistics the line width was fixed
at 64 eV as obtained from the phase-averaged spectra. 

To verify if the hardness ratio 
variations can be caused by changes in the absorption line parameters we fit the column density
simultaneously as a single free parameter to all sixteen spectra. The energy of the line
(kept equal for the spectra from the same phase interval) and the black-body temperature 
(for all spectra) were allowed to be fit individually. 
The resulting fit parameters are summarized in Table~\ref{tab-phase-fits}.
The spectra from ``soft'' and ``hard'' phases (as defined in Table~\ref{tab-phase-fits})
are presented in Fig.~\ref{tab-phase-fits}. No significant shift of the 
line energy between the two phase intervals is seen. The depth of the line varies 
strongly, which results in an \eqw\ during the ``soft" phase (around intensity maximum)
about a factor of two smaller than that during the ``hard" phase (at the end of the 
declining part of the pulse). At the same time, there is a 
2-3 eV increase in the black-body temperature found during the intensity maximum. 
At this phase the softening due to reduced line absorption dominates the spectral 
hardening caused by 
increased temperature and yields the hardness ratio minimum. 

To see if the variations in the line depth are really required to account for the 
spectral changes, we forced the line depth to be a single fit parameter, kept the same for all four
phase intervals. This results in a fit with $\chi^2$/dof = 2394/1970 in comparison to 
2267/1967 with free variation in line depth (Table~\ref{tab-phase-fits}). 
Comparison of the two models using an F-test yields a probability of 4.3\expo{-23} that the line depth 
is not variable. Allowing variations in column density \nh\ 
instead of variable absorption line (depth and energy are constant with phase) 
yields $\chi^2$/dof = 2336/1970 with
reduced \nh\ during ``soft" phase as was already reported by C01.
Also for this case the F-test probability of 9.7\expo{-13} strongly favors the model with 
variable line depth.
Our new results show that the model with phase-dependent absorption line yields the best 
description of the phase-resolved spectra of \rxb. 

\begin{table}[t]
\caption[]{Pulse-phase resolved spectra.}
\begin{tabular}{lccccc}
\hline\noalign{\smallskip}
\multicolumn{6}{c}{Model B: phabs*(bbody+gaussian); $\chi^2$/dof = 2267/1967} \\
\noalign{\smallskip}\hline\noalign{\smallskip}

\multicolumn{1}{l}{Rev.} &
\multicolumn{1}{c}{kT} &
\multicolumn{1}{c}{\nh} &
\multicolumn{1}{c}{\eline} &
\multicolumn{1}{c}{\wline} &
\multicolumn{1}{c}{\eqw} \\

\multicolumn{1}{l}{} &
\multicolumn{1}{c}{[eV]} &
\multicolumn{1}{c}{[\oexpo{20}cm$^{-2}$]} &
\multicolumn{1}{c}{[eV]} &
\multicolumn{1}{c}{[eV]} &
\multicolumn{1}{c}{[eV]} \\

\noalign{\smallskip}\hline\noalign{\smallskip}
\multicolumn{6}{l}{pulse minimum}\\
078 & 83.0$\pm$0.6 & 0.88$\pm$0.10 & 285$\pm$6 & 64f & -46.1$\pm$3.9 \\
175 & 82.4$\pm$0.7 & =1$^{(1)}$    & =1        &  =1 & -46.0$\pm$3.9 \\
533 & 86.5$\pm$0.7 & =1 	   & =1        &  =1 & -47.1$\pm$3.9 \\
534 & 86.2$\pm$0.7 & =1 	   & =1        &  =1 & -47.0$\pm$3.9 \\
\noalign{\smallskip}\hline\noalign{\smallskip}
\multicolumn{6}{l}{pulse rise}\\
078 & 84.2$\pm$0.6 & =1 	   & 288$\pm$8 &  =1 & -31.6$\pm$4.2 \\
175 & 83.3$\pm$0.7 & =1 	   & =5        &  =1 & -31.5$\pm$4.2 \\
533 & 88.0$\pm$0.7 & =1 	   & =5        &  =1 & -32.4$\pm$4.2 \\
534 & 87.0$\pm$0.7 & =1 	   & =5        &  =1 & -32.2$\pm$4.2 \\
\noalign{\smallskip}\hline\noalign{\smallskip}
\multicolumn{6}{l}{pulse maximum, ``soft" phase}\\
078 & 84.2$\pm$0.6 & =1 	   & 271$\pm$9 &  =1 & -31.1$\pm$4.4 \\
175 & 83.2$\pm$0.7 & =1 	   & =9        &  =1 & -30.9$\pm$4.4 \\
533 & 87.2$\pm$0.7 & =1 	   & =9        &  =1 & -31.9$\pm$4.4 \\
534 & 87.2$\pm$0.7 & =1 	   & =9        &  =1 & -31.9$\pm$4.4 \\
\noalign{\smallskip}\hline\noalign{\smallskip}
\multicolumn{6}{l}{pulse decline, ``hard" phase}\\
078 & 82.4$\pm$0.5 & =1            & 279$\pm$5 &  =1 & -57.3$\pm$3.5 \\
175 & 81.4$\pm$0.6 & =1            & =13       &  =1 & -57.0$\pm$3.5 \\
533 & 84.9$\pm$0.6 & =1 	   & =13       &  =1 & -58.3$\pm$3.5 \\
534 & 84.9$\pm$0.6 & =1 	   & =13       &  =1 & -58.3$\pm$3.5 \\
\noalign{\smallskip}\hline\noalign{\smallskip}
\end{tabular}

$^{(1)}$ "=n" denotes fit parameter is linked with parameter in line n
\label{tab-phase-fits}
\end{table}

\begin{figure}
\begin{center}
\resizebox{8.6cm}{!}{\includegraphics[clip=]{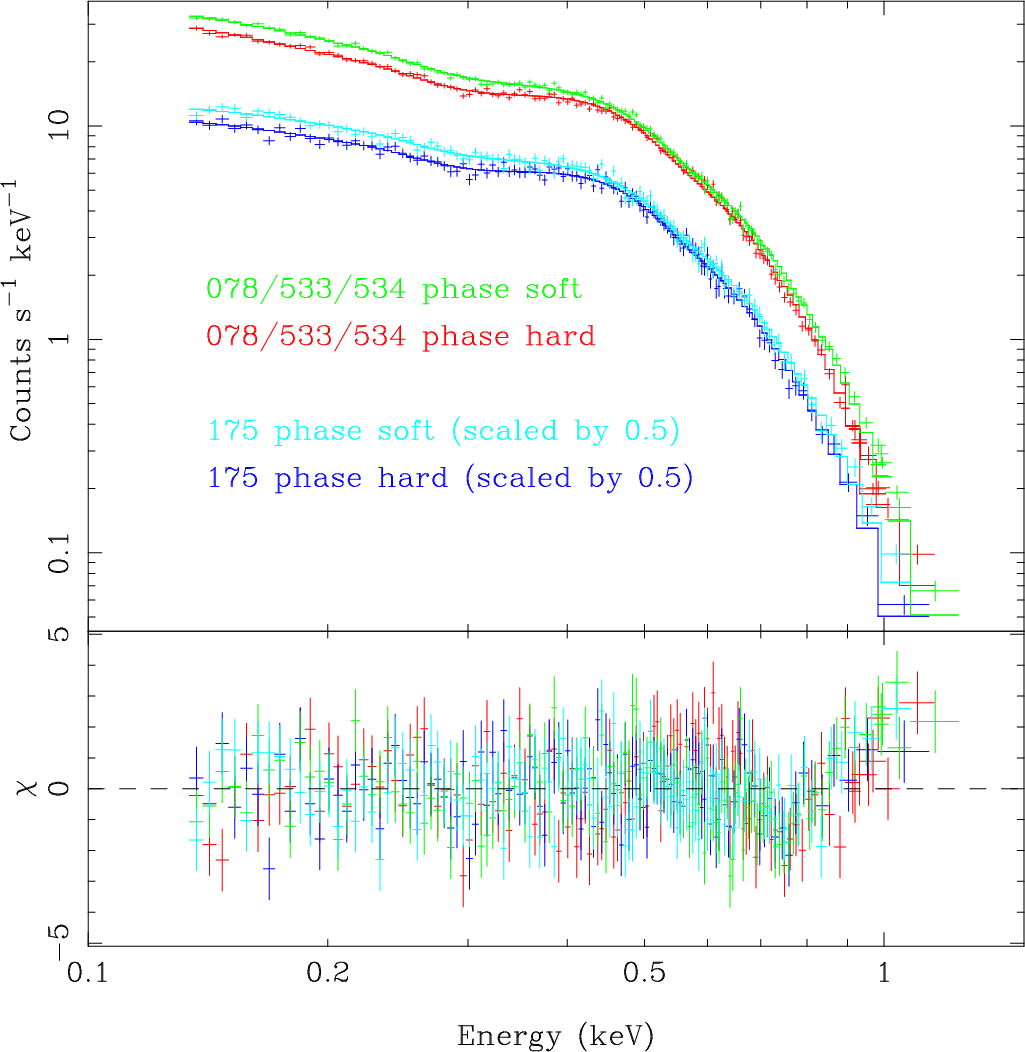}}
\end{center}
\caption{Phase-resolved EPIC-pn spectra from phases of high and low hardness ratios (see 
Fig.~{fig-hardness}). The upper pair of spectra shows the combined data from the three 
thin filter observations, the lower pair corresponds to the medium filter observation.
The upper and lower spectra in each pair are extracted from the phases of low (``soft" phase)
and high (``hard" phase) hardness ratio.}
\label{fig-phase-spectra}
\end{figure}

\section{Discussion}

From a thorough spectral analysis of four XMM-Newton observations of the isolated
neutron star \rxb\ we find deviations in the spectra from a Planckian shape, similar to those
reported by H03 for \rbs. A model consisting of a black-body continuum attenuated
by photo-electric absorption and an absorption line yields acceptable fits to
the XMM-Newton spectra and is also consistent with the Chandra LETGS spectrum currently
available from \rxb. From the pulse-phase averaged spectra we derive a line energy of
271 eV, $\sigma$ = 64 eV and \eqw\ = $-$40 eV, for a Gaussian-shaped line.
These absolute values may still be subject to systematic uncertainties larger than the 
statistical errors due to cross-calibration problems of the different instruments used for our 
analysis, but relative changes of model parameters, e.g. with pulse phase, could be determined
to high accuracy.

Hardness ratio variations with the pulse phase from \rxb\ were already reported by C01, 
indicating spectral variations. When formally fit with an absorbed 
black-body model the spectrum near intensity maximum with lower hardness ratio yields
a reduced \nh\ compared to phases around hardness ratio maximum. C01 suggested that this
could be explained by energy-dependent beaming effects. Alternatively, they discuss
variable absorption by matter captured by and rotating with the magnetosphere of the 
neutron star. Taking into account the absorption line revealed in the phase-averaged
spectra we find that the modulation of the hardness ratio with pulse phase can be better 
explained by variations in the absorption line depth which lead to changes in the line
\eqw\ by about a factor of two. During pulse rise and at intensity maximum, 
where we see a hardness ratio minimum, the absorption line is weakest. 
The black-body temperature varies by 2-3 eV between ``hard" and ``soft" phase spectra, 
with the highest temperature
during intensity maximum. The resulting hardness ratio minimum shows that the 
decrease of the absorption line depth is the dominating effect near pulse maximum.

H03 suggested cyclotron resonance absorption as a likely origin for the absorption 
feature seen in the X-ray spectra of \rbs. Since charged particles as electrons, protons or
ions can be the origin of this absorption one cannot uniquely derive the magnetic field 
strength from the cyclotron line energy without additional independent information. First
estimates for the magnetic field strength of \rxb\ were inferred from measurements of the
spin down rate under the assumption of magnetic dipole braking. 
\citet{2002MNRAS.334..345Z} give a
magnetic field B$\sim$\oexpo{13} G, which would exclude electrons as the origin of the
cyclotron line, which is expected at energies around 100 keV for such a strong field. 
In the case of protons a cyclotron line energy of 271 eV 
would result in a magnetic field strength of 5.3\expo{13} G for a neutron star with canonical
mass and radius (1.4 M$_{\sun}$, 10 km). Highly ionized atoms of heavy elements have
a mass to charge ratio of $\sim$2 with respect to protons and would lead to B a factor of
$\sim$2 higher than that derived for protons. Different ionization states would result in a
series of lines with energies differing by only a few percent. Smearing out of these lines due
to the B field variations in a rotating dipole field may cause a single broad absorption 
feature, broader than that expected from the B field smearing 
\citep{1995A&A...297..441Z,2001ApJ...560..384Z} of a single line and hardly resolvable.
In a most recent re-analysis of the available archival and new XMM-Newton data of \rxb\ 
\citet{2004Cropper} constrain B to (2.8 - 4.2)\expo{13} G from the pulse period history 
by interpreting the spin-down rate in terms of magnetic dipole radiation losses. This 
result is most compatible with the cyclotron line scenario involving protons.

\begin{figure}
\begin{center}
\resizebox{8.6cm}{!}{\includegraphics[clip=,angle=-90]{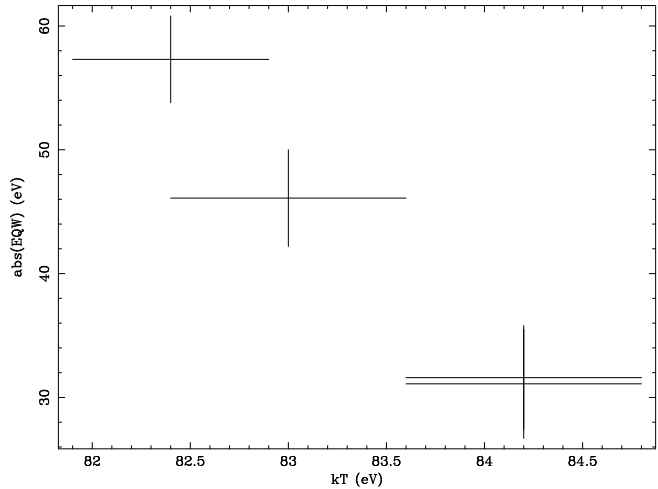}}
\end{center}
\caption{Black-body temperature vs. absorption line equivalent width derived from the EPIC-pn
spectra from the four pulse phase intervals described in the text and listed in
Table~\ref{tab-phase-fits} (only the values derived from the first observation are plotted).}
\label{fig-kt-eqw}
\end{figure}

An alternative possibility for the origin of the absorption line is atomic bound-bound 
transitions. In strong magnetic fields atomic orbitals are distorted into a cylindrical 
shape and the electron energy levels are similar to Landau states, with binding energies of atoms 
significantly increased. E.g. for hydrogen in a magnetic field of 
the order of \oexpo{13} G the strongest atomic transition is expected at energy 
E/eV $\approx$ 75(1+0.13ln(B$_{13}$))+63B$_{13}$,
with B$_{13}$ = B/\oexpo{13} G \citep{2002nsps.conf..263Z}. For the line energy found 
in the spectra of \rxb\ this would require B $\simeq$ 3\expo{13} G. Atomic line 
transitions are expected to be less prominent at higher
temperatures because of a higher ionization degree \citep{2002nsps.conf..263Z}. 
Qualitatively, this is in line with the observed anti-correlation between temperature 
and equivalent width (see Fig.~\ref{fig-kt-eqw}). However, the observed effect is much 
too strong: a temperature change of 2 eV at a temperature of 83 eV would not cause a 
change of the number of ionization states by a factor of two. In RBS 1223 the effect 
is even stronger as it shows an \eqw\ of --150 eV at similar temperature. At the same 
time in this source the pulsed fraction is much larger than in \rxb\, suggesting 
that the change of angles between the magnetic axis and the line of sight is much 
larger. We conclude that the anti-correlation between temperature and equivalent 
width does not reflect a direct physical connection, but is due to the fact that 
both correlating quantities are depending on the pulse phase.

Therefore, we prefer the interpretation of the observed line feature in terms of the proton 
cyclotron resonance. The observed dependence of temperature and equivalent width on pulse phase 
may be due to the change in the viewing geometry of the inclined magnetic rotator. 
However, a thorough quantitative modeling of cyclotron and atomic bound-bound 
absorption effects in strong magnetic fields (a few \oexpo{13} G) of isolated neutron stars 
and their observable rotational phase dependences is required to understand the behavior
in detail. 

\begin{acknowledgements}
The XMM-Newton project is supported by the Bundesministerium f\"ur Bildung und
For\-schung / Deutsches Zentrum f\"ur Luft- und Raumfahrt (BMBF / DLR), the
Max-Planck-Gesellschaft and the Heidenhain-Stif\-tung. We thank the anonymous referee
for valuable comments which helped to improve the paper.
\end{acknowledgements}

\bibliographystyle{apj}
\bibliography{ins,general,myrefereed,myunrefereed,mytechnical}

\end{document}